# Random Sequences Using the Divisor Pairs Function

Subhash Kak

**Abstract.** This paper investigates the randomness properties of a function of the divisor pairs of a natural number. This function, the antecedents of which go to very ancient times, has randomness properties that can find applications in scrambling, key distribution, and other problems of cryptography. It is shown that the function is aperiodic and it has excellent autocorrelation properties.

## INTRODUCTION

One of the earliest well-articulated mathematical problems to be found in the world literature is that of divisors of a number [1]. The problem of divisor pairs of a number is mentioned in the second millennium BCE Sanskrit text called the Śatapatha Brāhmaṇa[1]. This scientific knowledge described in the text includes geometry, algebra, and astronomy and it is one of the earliest sources for Pythagorean triples [2]. The idea of divisors in the text is in the sense that 4 has three divisors (1,2,4) and two divisor pairs (1,4) and (2,2).

Speaking of divisors of a number, the text states that the number 720 has 15 divisor pairs and 10800 has 30 divisor pairs. These numbers are correct since the number of divisors, $d(n)$, for $n = p_1^{a_1} \times p_2^{a_2} \times p_3^{a_3}$, equals $(a_1+1)(a_2+1)(a_3+1)$, and the number of divisor pairs which we call $\delta(n)$, will be half of it but rounded up when $d(n)$ is odd. In general,

$$\delta(n) = \lceil d(n)/2 \rceil$$

For the first example from the Śatapatha Brāhmaṇa, $720 = 2^4 \times 3^2 \times 5^1$, so $d(720) = 5 \times 3 \times 2 = 30$ and $\delta(720) = 15$. Similarly, $\delta(10800) = 30$. Parenthetically, it should be mentioned that these numbers are mentioned for their significance as the number of days in the half-month and the month [3].

The number $\delta(n)$ is related to $v(n)$, the *valency* of the number $n = p_1^{a_1} p_2^{a_2} p_3^{a_3} ... p_n^{a_n}$, that is defined to be $v(n) = a_1 + a_2 + a_3 + ... + a_n$. Figure 1 gives the value of the function $\delta(n)$ for n < 1000.

The function $v(n)$ satisfies the relation: $v(ab) = v(a) + v(b)$. Below is a list of beginning natural numbers with valency of 1, 2, 3, and so on:

$v(n) = 1$:      1, 2, 3, 5, 7, 11, 13, … (prime numbers)
$v(n) = 2$:      4, 6, 9, 10, 14, 15, … (squares and semi-primes)
$v(n) = 3$:      8, 12, 18, 20, 27, …
$v(n) = 4$:      16, 24, 36, 40, 54, …

Functions of $v(n)$ are well known in the mathematics literature [4],[5]. Related also to the valence number is the number of primes factors $b(n)$, which is simply equal to *n*.

---

[1] Some scholars who do not consider the internal astronomical evidence in the text date it to the first half of first millennium BCE.

Another related function is $\kappa(n) = (-1)^{\delta(n)}$ for the sequence of numbers n. We also define $S(n) = \sum_{1}^{n} \kappa(i)$, which is the running sum of the $\kappa(n)$ function. Table 1 provides the values of $v(n)$, $d(n)$, $\delta(n)$, $\kappa(n)$, and $S(n)$ for n ≤ 16.

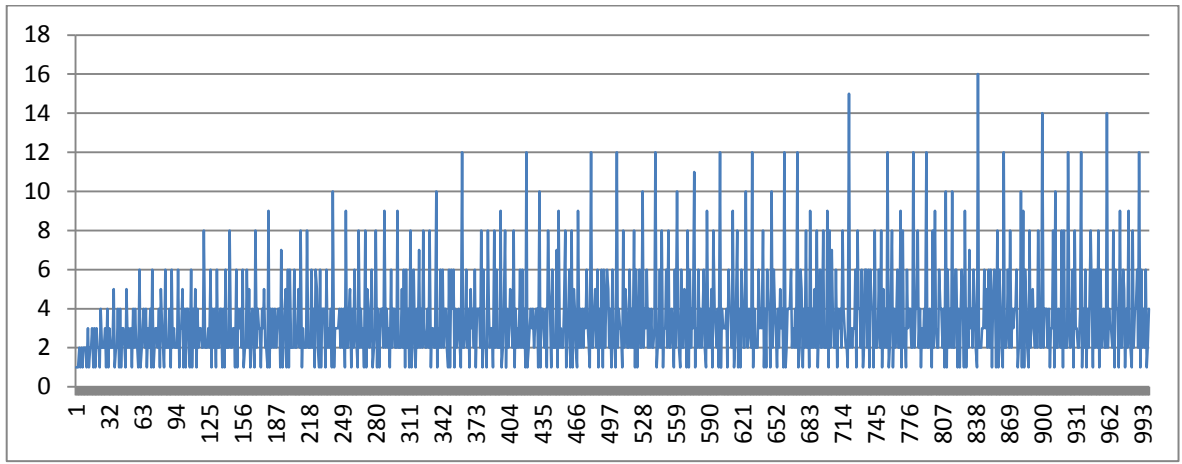

**Figure 1.** The function $\delta(n)$ for n < 1,000

**Table 1.** Values of $v(n)$, $d(n)$, $\delta(n)$, $\kappa(n)$, and $S(n)$

| n | 1 | 2 | 3 | 4 | 5 | 6 | 7 | 8 | 9 | 10 | 11 | 12 | 13 | 14 | 15 | 16 |
|---|---|---|---|---|---|---|---|---|---|----|----|----|----|----|----|----|
| $v(n)$ | 1 | 1 | 1 | 2 | 1 | 2 | 1 | 3 | 2 | 2 | 1 | 3 | 1 | 2 | 2 | 4 |
| $d(n)$ | 1 | 2 | 2 | 3 | 2 | 4 | 2 | 4 | 3 | 4 | 2 | 6 | 2 | 4 | 4 | 5 |
| $\delta(n)$ | 1 | 1 | 1 | 2 | 1 | 2 | 1 | 2 | 2 | 2 | 1 | 3 | 1 | 2 | 2 | 3 |
| $\kappa(n)$ | -1 | -1 | -1 | 1 | -1 | 1 | -1 | 1 | 1 | 1 | -1 | -1 | -1 | 1 | 1 | -1 |
| $S(n)$ | -1 | -2 | -3 | -2 | -3 | -2 | -3 | -2 | -1 | 0 | -1 | -2 | -3 | -2 | -1 | 0 |

In general, we can consider other functions of the exponents associated with the prime factors of a number. We can thus speak of a generalized valence function, $V(n)$, which is given by:

$$V(n) = f(a_1, a_2, ..., a_n)$$

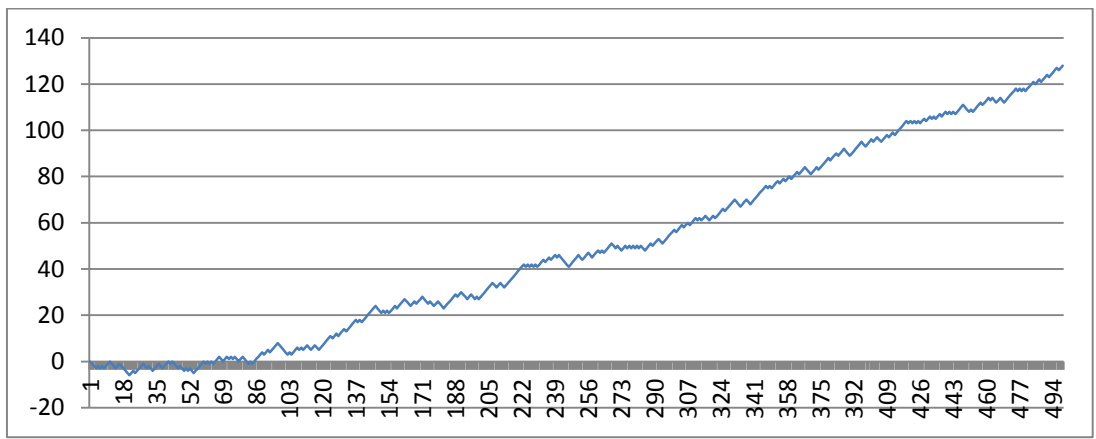

**Figure 2.** The function $S(n)$ for n < 500

Thus $\delta(n)$ is a special case of $V(n)$. In general, it would be worthwhile to determine which forms of $V(n)$ are of most interest to the computer scientist from the point of view of generating random sequences.

Here we wish to study the randomness properties of the binary sequence $\kappa(n) = (-1)^{\delta(n)}$. This is continuation of a project to examine the randomness characteristics of a variety of number-theoretic functions which include prime reciprocals [6]-[8], Pythagorean triples [9], permutation transformations [10], and Goldbach sequences [11]. We show that the sequence is irrational and it can be used as a pseudorandom sequence.

## A RELATED FUNCTION

The Liouville function $\lambda(n)$ is a binary function of $v(n)$ that maps even values to 1 and odd values to -1:

$$\lambda(n) = (-1)^{v(n)}$$

Table 2 presents a comparison of the values of $\kappa(n)$ and $\lambda(n)$ for $n \leq 16$.

**Table 2.** Comparison of $\kappa(n)$ and $\lambda(n)$

| $n$ | 1 | 2 | 3 | 4 | 5 | 6 | 7 | 8 | 9 | 10 | 11 | 12 | 13 | 14 | 15 | 16 |
|---|---|---|---|---|---|---|---|---|---|---|---|---|---|---|---|---|
| $\kappa(n)$ | -1 | -1 | -1 | 1 | -1 | 1 | -1 | 1 | 1 | 1 | -1 | -1 | -1 | 1 | 1 | -1 |
| $\lambda(n)$ | 1 | -1 | -1 | 1 | -1 | 1 | -1 | -1 | 1 | 1 | -1 | -1 | -1 | 1 | 1 | 1 |

As seen in Table 2, $\kappa(n)$ and $\lambda(n)$ are different at values of $n = 8, 16$. Likewise, the values will be different for n= 27, 81, and so on.

The Liouville function $\lambda(n)$ satisfies the following property:

$$\sum_{d|n} \lambda(d) = \begin{cases} 1, & n = perfect\ square \\ 0, & otherwise \end{cases}$$

A number with high valency is composite in multiple ways. It is interesting that Ramanujan worked on highly composite numbers [12] and this work has attracted recent attention [13]. An integer $n$ is said largely composite if for m $\leq$ n, $d(m) \leq d(n)$.

Ramanujan [12] presented the following result relevant to *n*. If $\sigma_{-s}(N)$ denotes the sum of the inverses of the *s*th powers of the divisors of *n*, then

$$\sigma_{-s}(N) < \frac{\{1-(p_1 p_2 p_3 ... p_n N)^{-s/n}\}^n}{(1-p_1^{-s})(1-p_2^{-s})...(1-p_n^{-s})}$$

For *s*=0, $\sigma_0(N) = d(N)$ is the number of divisors of *N*.

# THE FUNCTIONS $\delta(n)$ AND $\kappa(n)$

The properties of the divisor pairs function $\delta(n)$ are obviously derivable from that of the divisor function $d(n)$. These two functions satisfy the following properties:

$$d(n_1 n_2) = d(n_1)d(n_2) \text{ if } \gcd(n_1, n_2) = 1$$
$$d(n_1 n_2) \leq d(n_1)d(n_2) \text{ if } \gcd(n_1, n_2) \neq 1$$

$$\delta(n_1 n_2) < 2\delta(n_1)\delta(n_2) \text{ if } \gcd(n_1, n_2) \neq 1$$
$$\delta(n_1 n_2) \leq 2\delta(n_1)\delta(n_2) \text{ if } \gcd(n_1, n_2) = 1$$
$$\delta(n_1 n_2) = 2\delta(n_1)\delta(n_2) \text{ if } \gcd(n_1, n_2) = 1 \text{ and both } d(n_1) \text{ and } d(n_2) \text{ are even.}$$

For the function $\kappa(n)$ the following properties are evident (where $p$ is prime):

$$\kappa(p) = -1$$
$$\kappa(p^2) = 1$$
$$\kappa(p^n) = \begin{cases} -1, & n = 1,4,5,8,9,12... \\ 1, & n = 2,3,6,7,10,11,... \end{cases}$$
$$\kappa(p_1 p_2) = 1$$

Figure 3 presents the values of the function $\kappa(n)$ for $n < 100$.

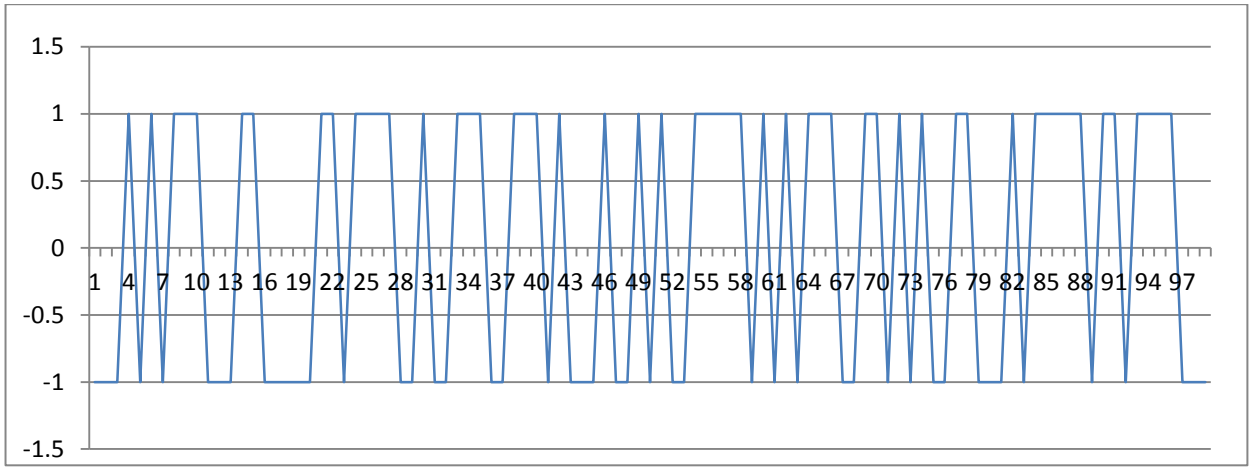

**Figure 3.** The function $\kappa(n)$ for $n < 100$

**Theorem.** *There is no periodicity associated with the $\kappa(n)$ function for any $n > N$.*

*Proof.* We establish this result by assuming it is true and then showing that leads to a contradiction. Let the $\kappa(n)$ sequence have a period of $k$ after $n=N$. This would imply that $\kappa(N) = \kappa(N+k)$ and $\kappa(N+r) = \kappa(N+r+k)$ for any $r$. Choose $N+r = p$ so that $(p+k) = p_1 p_2$. This condition that $k = p_1 p_2 - p$ for a random $p$, which is that a number may be written as a semiprime minus a prime, is true from experimental calculations and also for large numbers [14]. This would imply that $\kappa(p) = \kappa(p_1)\kappa(p_2)$ or $\kappa(p) = \kappa(p_1^2)$, which is impossible.

This theorem establishes that $\kappa(n)$ is an irrational function and, therefore, it can be used as a pseudorandom sequence [15].

The autocorrelation function captures the correlation of data with itself. For a data sequence $a(n)$ of $N$ points the autocorrelation function $C(k)$ is represented by

$$C(k) = \frac{1}{N}\sum_{j=1}^{N} a(j)a(j+k)$$

For a noise sequence, the autocorrelation function $C(k) = E(a(i)a(i+k))$ is two-valued, with value of 1 for k=0 and a value approaching zero for k≠0 for a zero-mean random variable. Since $S(n)$ drifts towards increasing positive values, for any choice of $N$, it would have a non-zero mean $\mu$ associated with it. Assuming ergodicity, such a sequence will have C(k) as 1 for k=0 and approximately $\mu^2$ for non-zero k.

Figures 4 and 5 present the autocorrelation function of the series $\kappa(n)$ for n =1000 and 5000, respectively. The value of $\mu_{1000}$=0.326 and $\mu_{5000}$=0.462. The value of $C(k)$ for non-zero k is therefore centered around 0.106 and 0.213, respectively.

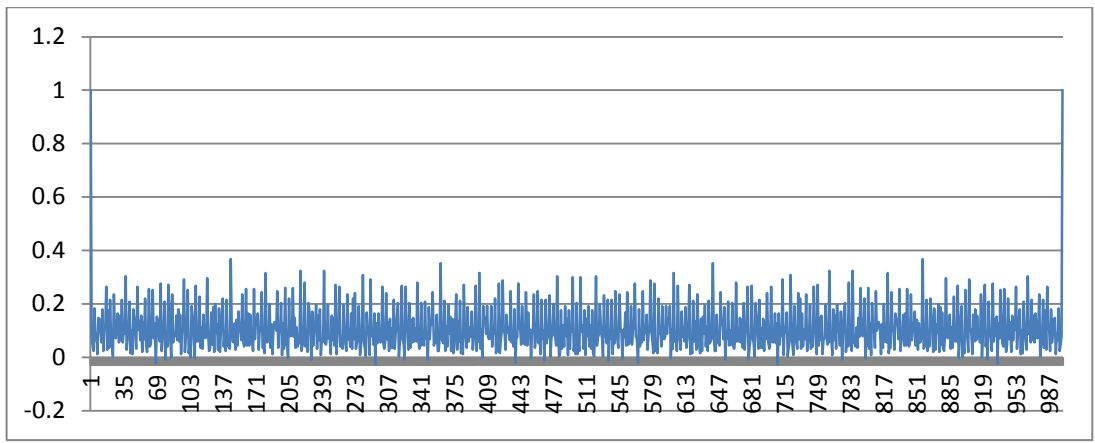

**Figure 4.** The autocorrelation function for $\kappa(n)$ for n =1000

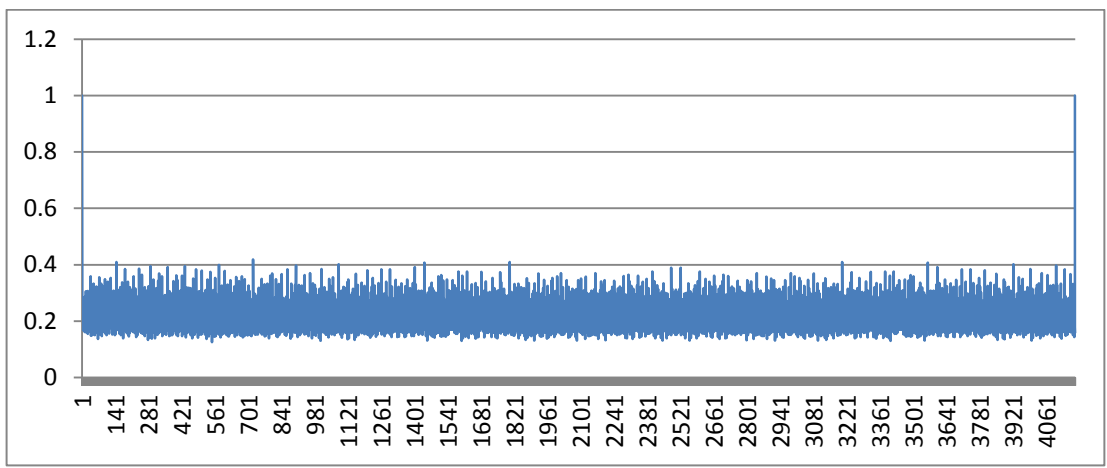

**Figure 5.** The autocorrelation function for $\kappa(n)$ for n =5000

As N becomes large the variance of the values in the autocorrelation function will reduce and in the limit it will be zero.

## CONCLUSIONS

This paper examined the properties of the divisor pairs function. In particular, the binary sequence $\kappa(n) = (-1)^{\delta(n)}$, which is closely related to the Liouville function, was investigated for its randomness characteristics. While its running sum drifts to positive values, its autocorrelation function is approximately two-valued which means that it can find applications in many cryptography applications. Many interesting questions remain: These include behavior of $S(n)$ for large values of *n* and the use of other functions of the valency of a number to generate random sequences.

**Acknowledgement.** This research was supported in part by research grant #1117068 from the National Science Foundation.